\documentclass[11pt,oneside,letterpaper]{article}
\pdfoutput=1

\usepackage{setspace}
\usepackage{graphicx}
\usepackage{subcaption}
\usepackage{amsmath}
\usepackage{amssymb}
\numberwithin{equation}{section}
\usepackage{physics}
\usepackage{xcolor}
\usepackage{color}  \definecolor{darkgreen}{rgb}{0,0.5,0}
\definecolor{darkred}{rgb}{0.5,0,0}    \definecolor{darkblue}{rgb}{0,0,0.6}
\definecolor{purple}{rgb}{0.4,.2,0.7}
\definecolor{black}{rgb}{0,0,0}
\usepackage[hyperfootnotes = true, colorlinks = true, linkcolor = darkred, citecolor = darkred, urlcolor = darkred]{hyperref}
\usepackage{comment}
\usepackage{csquotes}
\usepackage{cite}
\usepackage{float}
\usepackage{units}
\usepackage{tikz}
\usepackage{cancel}
\usepackage{mathtools}
\usepackage{pdfsync}
\usepackage{slashed}
\usepackage{url}
\usepackage{upgreek}
\usepackage{float}
\usepackage{units}
\usepackage{tikz}
\usepackage{tikz-feynman}
\usepackage{cancel}
\usepackage{orcidlink}
\usepackage[margin = 2.2cm]{geometry}
\usepackage[ragged]{footmisc}
\newcommand{\be}{\begin{eqnarray}\displaystyle}
\newcommand{\ee}{\end{eqnarray}}
\newcommand{\nn}{\nonumber}
\newcommand{\f}{\frac}
\newcommand{\p}{\partial}

\newcommand{\gym}{g_{_\text{YM}}}

\begin{document}
\include{feynmandefn}
\thispagestyle{empty}
\begin{center}
    ~\vspace{5mm}
    
    {\Large \bf 
        Subleading Chern--Simons soft factors in perturbative de Sitter
    }
    
    \vspace{0.4in}

    {\bf Pratik Chattopadhyay$^{1,}$$^2$\orcidlink{0000-0002-5131-6334}, Avi Wadhwa$^3$\orcidlink{0000-0002-6724-0872}}

    \vspace{0.4in}
    
    $^1$ Tsung-Dao Lee Institute, Shanghai Jiao Tong University, \\1 Lisuo Road, Pudong New Area, Shanghai 201210, China.
    
    \vspace{0.4in}
    
    $^2$ School of Physics, The University of Electronic Science and Technology of China,\\ No.2006, Xiyuan Avenue, West Hi-Tech Zone, Chengdu, Sichuan, P.R.China, Post Code: 611731
    
    \vspace{0.4in}
    
    $^3$ International Centre for Theoretical Sciences,\\ Tata Institute of Fundamental Research,\\Shivakote,  Hesaraghatta Hobli,
    Bengaluru 560089, Karnataka,  India.
    
    \vspace{0.1in}
    
    {$^{1,2}$ \tt pratikpc@gmail.com}, {$^3$ \tt avi.wadhwa@icts.res.in}
    
    \vspace{0.4in}
    
\end{center}

\begin{abstract}

Chern--Simons perturbations introduce corrections to soft theorems for gauge theories at subleading $\mathcal{O}\left(\omega^0\right)$ order in soft momenta. We investigate these soft theorems in flat spacetime with perturbative $1/\ell^2$ de Sitter corrections. Following previous works, we define the perturbative scattering matrix in a compact region in the static patch of de Sitter. We start with computing the soft theorem for an $n$-gluon amplitude in pure Yang--Mills. We then add a Chern--Simons perturbation and compute the correction to the subleading soft theorem. We show that Chern--Simons corrections do not mix with the $1/\ell^2$ de Sitter curvature corrections at subleading order $\mathcal{O}\left(\omega^0\right)$. Alternatively, one can say that the subleading Chern--Simons soft factors are insensitive to the de Sitter curvature at this order, indicating their topological nature at the level of amplitudes. This also suggests a universal behavior of these Chern--Simons soft factors.

\end{abstract}

\pagebreak

\tableofcontents

\section{Introduction}

The study of the infrared structure of scattering amplitudes of gauge theories has been of considerable interest throughout the years. An important aspect of these gauge theory amplitudes is that they show certain universal factorisation properties when one of the external gauge bosons becomes soft-- i.e. their energy is very low compared to the rest of the external particles. These properties are encoded into what are known as soft theorems \cite{Bloch_1937, Gell-Mann_Goldberger, Low_1954, Low_1958,  Weinberg_1965}. These universal soft factors have also been realised as Ward identities for the asymptotic symmetries of the underlying theory \cite{Temple_He, Laddha_2015, Daniel_Kapec, Lysov_Pasterski_Strominger}. Soft theorems have also been related to memory effects \cite{Strominger_Zhiboedov, Pasterski_Strominger_Zhiboedov, Strominger}.

There has been emerging interest to understand soft theorems and their connections with the asymptotic symmetries in curved spacetime. Significant progress have been made in the context of anti-de Sitter space \cite{Fernandes_review_2023, Banerjee_2021, Banerjee_Bhattacharjee_2021, Banerjee_2023}. However, since both inflation and late time acceleration of the universe is governed by de Sitter spacetime, it is important to understand these theorems in such a setting. Recent work has shown that in the small cosmological constant limit, the leading soft factor undergoes perturbative corrections due to the background de Sitter geometry \cite{Sayali_Diksha, SCB, PD, Pratik}. In addition, the soft theorems in the full static patch of de Sitter space have also been derived from the Ward identities of the near-horizon symmetries \cite{Mao_Zhou, Mao_Zhang}.

The leading and subleading soft theorems have also been understood as an artifact of local on-shell gauge-invariance of the underlying theory \cite{Bern, Sen_2017, SenII_2017, Laddha_Sen_2017, Chakrabarti_Sen_2017, Sayali_Sahoo_2018}. Recently, it was shown that the subleading soft factor gets corrections in the presence of non-trivial gauge invariant perturbations, like a Chern--Simons deformation in $D=5$ \cite{Avi}. Chern--Simons terms \cite{Chern:1974ft} in the action are gauge-invariant only up to a boundary term. Therefore, the gauge invariance is only under small gauge transformations which vanish as we move towards the boundary. This indicates that the subleading soft factor is not quite as universal as was previously assumed, but gets corrected with Chern--Simons soft factors in $D=5$.

Chern--Simons terms can only be defined in odd spacetime dimensions. They are perturbative to Yang--Mills theory in $D=5$ or higher\footnote{In $D=3$, the Chern--Simons terms are more relevant in IR than the Yang--Mills terms.}. As argued in \cite{Avi} they correct the subleading soft theorem only in $D=5$ for gluons and photons. In higher dimensions (say $D=7$), there are no 3-point Chern--Simons vertices to contribute at the subleading order. Also, there are no pure gravitational Chern--Simons terms in $D=5$\footnote{Gravitational Chern--Simons terms only exist in 4p-1 spacetime dimensions (for positive integer p) while the gauge Chern--Simons terms may exist in any odd spacetime dimensions.} and therefore Chern--Simons terms do not correct the gravitational soft theorems.

Chern--Simons terms in the action are also topological-- they are independent of the background metric. Therefore one may ask whether this translates to a statement at the level of amplitudes. This motivates us to compute $1/\ell^2$ de Sitter corrections to these Chern--Simons soft factors, if any. In this work, we show that the subleading Chern--Simons soft factors in fact do not receive additional $1/\ell^2$ curvature corrections in the de Sitter background and retain the same form as in flat spacetime. This indicates that the subleading Chern--Simons soft factors are completely insensitive to curvature, as may be expected from their topological nature. 

All the computations are valid for tree level and no implications are made for higher loop orders.

The paper is organised as follows-- we describe the problem statement and key assumptions in \S\ref{sec:setup}. In \S\ref{sec:gf}, we describe the gluon field in de Sitter with the mode expansion and propagator. In \S\ref{sec:soft-gluon-theorem} we derive the subleading soft gluon theorem with de Sitter corrections in $D=5$ without any Chern--Simons terms. For simplicity of demonstration, we take the example of an $n$--gluon amplitude in pure Yang--Mills theory without any matter. Then in \S\ref{sec:CS} we follow the same steps to compute the contribution due to the addition of a Chern--Simons term (to a generic Yang--Mills theory with matter) as in \cite{Avi} but this time in flat spacetime with perturbative de Sitter $1/\ell^2$ corrections. We show that the subleading term does not get corrected at all orders in $1/\ell^2$.  This is non trivial since even though the Chern--Simons term in the action (and hence the corresponding vertex) is independent of the metric, the amplitude also contains metric-dependent external modes, propagators and LSZ factors.

\section{Setup}\label{sec:setup}
\subsection{De Sitter spacetime}
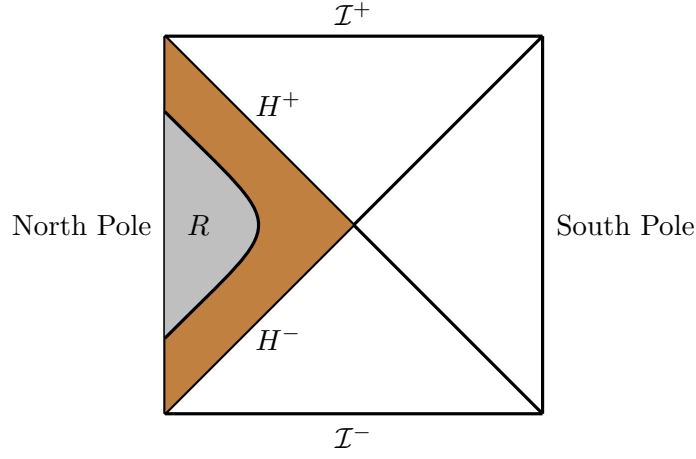
\begin{figure}
\centering
\begin{tikzpicture}
\draw[black, very thick] (-2.5,2.5) to node[shift={(0,0.3)}, sloped] {$\mathcal{I}^+$} (2.5,2.5);
\draw[black, very thick] (-2.5,-2.5) to node[shift={(0,-0.3)}, sloped] {$\mathcal{I}^-$} (2.5,-2.5);
\draw[black, very thick] (-2.5,2.5) to node[shift={(-1.1,0)}] {North Pole} (-2.5,-2.5);
\draw[black, very thick] (2.5,2.5) to node[shift={(1.1,0)}] {South Pole} (2.5,-2.5);
\draw[black, very thick] (-2.5,2.5) to node[shift={(-1,1.6)}] {$H^+$} (2.5,-2.5);
\draw[black, very thick] (2.5,2.5) to node[shift={(-1,-1.5)}] {$H^-$} (-2.5,-2.5);
\draw[fill=brown] (-2.5,2.5) -- (0,0) -- (-2.5,-2.5) -- cycle;
\draw[black, very thick,fill=lightgray] (-2.5,1.5) to [out=-45, in=45,looseness=2] node[shift={(-0.8,0)}] {$R$} (-2.5,-1.5);
\end{tikzpicture}
\caption{Penrose diagram of de Sitter space.}
\label{fig:dS_Penrose}
\end{figure}

We study gluon scattering with the emission of a soft gluon in the static patch of de Sitter spacetime. We confine this scattering process to the small compact region $R$ inside the static patch as shown\footnote{Figure \ref{fig:dS_Penrose} is adapted from \cite{Sayali_Diksha} with minor edits.} in fig. \ref{fig:dS_Penrose}. The de Sitter metric can be put in the conformally flat form in the stereographic coordinates $x^\mu$ as is shown in \cite{ Sayali_Diksha, SCB}, i.e,
\begin{equation}
g_{\mu\nu}=\Omega^2\eta_{\mu\nu}, ~~\Omega=\frac{1}{1+x^2/4\ell^2},
\end{equation}
where $\eta_{\mu\nu}$ is the Minkowski metric and $\ell=H^{-1}$ is the de Sitter length scale. Inside the compact region, $x^\mu\ll \ell$, we can expand the de Sitter metric up to $\mathcal{O}(\ell^{-2})$ about thef lat metric as follows:
\begin{equation}\label{eq: metric_expand}
g_{\mu\nu}\approx\Big(1-\frac{x^2}{2\ell^2}\Big)\eta_{\mu\nu}.
\end{equation}
Thus, we effectively work in the large curvature length or small cosmological constant limit. We now discuss the key assumptions on our setup and various scales involved in the scattering process.
\subsection{Key assumptions in the setup}
\begin{itemize}
\item We confine our scattering process in the region $R$ of the static patch in de Sitter spacetime. This region is much smaller than the de Sitter length scale $\ell$ and therefore we treat $1/\ell$ as a perturbation parameter in which the full scattering amplitude is expanded. The interaction time scale is much smaller than the time required for the particles to reach the boundary. Thus, the particles are free from interactions at early and late times. To make this statement precise, one can define free particle states on early- and late-time Cauchy slices, which are the Hilbert spaces of the incoming and outgoing states, respectively.
\item Unlike in flat spacetime, we have three energy scales in de Sitter: the energy $E$ of the scattered hard particles, the energy $\omega$ of the scattered soft particles and the Hubble constant $H=1/\ell$. We are already working in the perturbative limit in de Sitter, where\footnote{This follows from the limit: $x^\mu/\ell\ll1$.} $E,\omega\gg1/\ell$. Furthermore, we have the soft limit $\omega\ll E$. Therefore the natural way to organise the expansion is to define a dimensionless parameter $\delta=\omega \ell$. The amplitude is then expanded in $1/\delta$ and then $\omega/E$. The $1/\delta$ terms then denote the de Sitter corrections to the soft expanded amplitude.
\item The amplitude $\Gamma_n$ for the scattering of n--gluons of momenta $p_1,...,p_n$ with an emission of a soft gluon of momentum $k\ (=\omega\hat{k})$ in the presence of a Chern--Simons term decomposes in the soft limit $\omega\to 0$ as follows:
\begin{equation}
\Gamma_{n}(\{p_1,...,p_{n-1}\},\omega\hat{k}) = \Big[\mathbb{S}(\{p_i\},\omega\hat{k})\ +\mathbb{S}^{(1)\,\text{CS}}({p_i},\omega\hat{k})\Big]  \Gamma_{n-1}(\{p_1,...,p_{n-1}\}),
\end{equation}
\begin{equation}
\mathbb{S}(\{p_i\},\omega\hat{k}) = \frac{1}{\omega}\mathbb{S}^{(0)}(\{p_i\},\omega\hat{k})+\mathbb{S}^{(1)}(\{p_i\},\omega\hat{k}) + \dfrac{1}{\delta^2} \mathbb{S'}^{(1)}(\{p_i\},\omega\hat{k}),
\end{equation}
where $\mathbb{S}^{(0)}$ is the leading gauge soft factor, $\mathbb{S}^{(1)}$ is the subleading gauge soft factor, $\mathbb{S'}^{(1)}$ is the de Sitter correction to the subleading gauge soft factor and
$\mathcal{S}^{(1)\text{CS}}$ is the CS subleading soft factor. The final result of our analysis is that $\mathbb{S}^{(1)\text{CS}}$ does not receive any $1/\delta$ curvature corrections.
\end{itemize}

\section{Perturbative $\mathcal{S}$-matrix}
\label{S-matrix_sec}
In this section, we define a perturbative $\mathcal{S}$-matrix in the de Sitter space and deduce the curved spacetime generalization of the LSZ formula. To formulate a convenient $\mathcal{S}$-matrix, we need to prescribe asymptotic states at the boundaries of spacetime. However, in de Sitter spacetime, there is no notion of a global observable boundary. The only observable region is the static patch. This patch is bounded by the cosmological horizons. We thus treat the cosmological horizons as physical boundaries of this spacetime to define the so called asymptotic states. The particles can reach the late time Cauchy slices as mentioned previously in a finite time $T$. This time is assumed to be much longer than the interaction time scale. In addition, all the lengths in our calculations are very small compared to the de Sitter radius $l$, i.e., $|x|\ll \ell$.

The $\mathcal{S}$-matrix describing the scattering of incoming gluons with momenta $p_i$ to the outgoing gluons with momenta $p_j$ followed by the emission of gluons with momenta $p_k$ is defined as 
\begin{equation}
\label{Gamma_general}
    \Gamma (\{p_j,p_k\},\{p_i\}) = \lim_{t\to T} \left\langle \mathcal{T} \prod_{j,k\in \text{out}} \sqrt{2E_{p_j}} a^{h_j}_{p_j}(t)\ \sqrt{2E_{p_k}} a^{h_k}_{p_k}(t) \prod_{i\in \text{in}} \sqrt{2E_{p_i}} a^{h_i\dagger}_{p_i}(-t) \right\rangle,
\end{equation}
where $\mathcal{T}$ represents time-ordering.
To deduce the LSZ formula, we rewrite the above expression as follows:
\begin{multline}
\label{Gamma_2}
    \Gamma (\{p_j,p_k\},\{p_i\}) = \lim_{t\to T} \bigg\langle \mathcal{T} \prod_{j,k\in \text{out}} \sqrt{2E_{p_j}} \big(a^{h_j}_{p_j}(t) - a^{h_j}_{p_j}(-t)\big) \sqrt{2E_{p_k}}\big(a^{h_k}_{p_k}(t)-a^{h_k}_{p_k}(-t)\big) \\
    \cdot \prod_{i \in \text{in}} \sqrt{2E_{p_i}} \big(a^{h_i\dagger}_{p_i}(-t)-a^{h_i\dagger}_{p_i}(t)\big) \bigg\rangle,
\end{multline}
where all the annihilation (creation) operators go to the right (left) side in the time order product and annihilate the ``in" (``out") -state, so that the above expression reduces to Eq. (\ref{Gamma_general}). As shown in the appendix, inverting the inner product expression for the gluon modes, one can deduce
\begin{align}
\label{s_inner_invert}
a^h_p(T)-a^h_p(-T)&=-i\int d^5 x \sqrt{-g(x)}\eta^{\mu \nu} f^{*ah}_{p\nu} \left(\left(1+ \f{x^2}{\ell^2}\right)\Box \ -\ \f{1}{2\ell^2}\ [x.\p +1]\right)A^a_\mu (x). \nn\\ 
\end{align}
Substituting (\ref{lszA}) into Eq. (\ref{Gamma_2}) yields
\begin{align}
\Gamma(\lbrace p_i,p_k\}, \{p_j \rbrace)&=\int \prod_{i,k\in \text{in}}\prod_{j\in \text{out}}\ [d^5x_i] [d^5y_j] [d^5z_k] \  f_{p_{i}\sigma}^{*\,a_{i},h_{i}}(x_{i})\eta^{\sigma\mu_{i}}\ \nn\\
& (-i)\mathcal{D}_{y_j}\ (-i)\mathcal{D}_{x_i}\  (-i)\mathcal{D}_{z_k} \langle \mathcal{T}\ A^{a_1}_{\mu_1}(z_k)A^{a_2}_{\mu_2}({x_i})\ldots A^{a_{n}}_{\mu_n}({y_j})\rangle .\label{LSZ}
\end{align} 
where the measure factors are given by:
\begin{equation}
[d^5x_i] [d^5y_j] [d^5z_k] 
 = d^5x_i\ d^5y_j\ d^5z_k \ \sqrt{2E_{p_i} (2E_{p_j})\ (2E_{p_k})} \ \sqrt{-g(x_i) \ (-g(y_j)) \ (-g(z_k))}. \nn 
\end{equation}

\section{Example: De Sitter corrections to the soft gluon theorem for pure Yang--Mills theory}\label{sec:soft-gluon-theorem}

\begin{figure}[h]
\begin{center}
\vskip .7 cm 
\includegraphics[scale=.4]{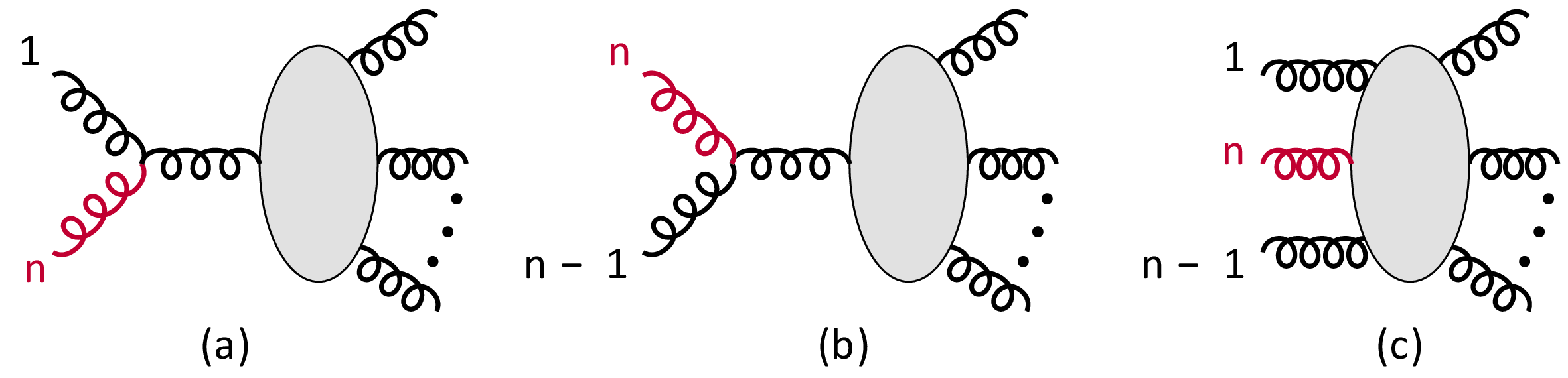}
\end{center}
\vskip -.7 cm 
\caption{(a,b) Gluon emitted from $1$st or $\left(n-1\right)$th external line. (c) Gluon emitted from an internal line}
\label{fig:soft_gluon}
\end{figure}

We closely follow the analysis done in \cite{Sayali_Diksha, SCB, Pratik}. As a demonstrative example, we consider an $n$--point colour ordered gluon amplitude $\Gamma_{n}(p_{1},\ldots,p_{n-1},k)$, where the $n$th gluon goes soft and hence $p_{n}=k$. The leading and subleading contributions come from two kinds of diagrams-- (a,b) where the soft gluon attaches to an external leg with momenta $p_{1}$ or $p_{n-1}$ \footnote{The soft gluon cannot attach to other external gluon lines without violating colour ordering.}, (c) where it attaches to an internal line as shown in fig. \ref{fig:soft_gluon} \footnote{Figure \ref{fig:soft_gluon} is adapted from \cite{Bern} with minor edits.},
\begin{align}
\mathcal{A}^{^\text{YM}}=\mathcal{A}^{^\text{YM}\left(1\right)}+\mathcal{A}^{^\text{YM}\left(n-1\right)}+\mathcal{A}^{^\text{YM}\left(\text{int}\right)}.
\end{align}
For calculating the (a,b) diagram, the term in the action corresponding to the 3--point vertex is given by,
\begin{equation}
\mathcal{S}_{3}=\int d^{5}x\,\mathcal{L}^{^\text{YM}}_3.
\end{equation}
with
\begin{equation}
\mathcal{L}^{^\text{YM}}_3=\sqrt{-g\left(x\right)}\;\,\gym\,f^{abc}\left(\partial_{\mu}A_{\nu}^{a}\right)g^{\rho\mu}g^{\sigma\nu}A_{\rho}^{b}A_{\sigma}^{c}
\end{equation}

In \cite{Sayali_Diksha, Pratik}, the derivation of the LSZ formula for photons was demonstrated in detail. The derivation remains identical for gluons apart from the colour factors. Thus we quote the final result, by modifying it accordingly for the gluons.
The LSZ formula for the $n$--gluon amplitude $\Gamma(p_{1},\ldots,p_{n})$  in de Sitter is given by,
\begin{equation}\label{eq:lsz}
\Gamma(p_{1},\ldots,p_{n})=\prod_{i=1}^{n}\left[\int \left[d^{5}x_{i}\right]\,\;f_{p_{i}\sigma}^{*\,a_{i},h_{i}}(x_{i})\eta^{\sigma\mu_{i}}\;(-i)\mathcal{D}_{x_{i}}\right]\;\big\langle \mathcal{T}A_{\mu_{1}}^{a_{1}}(x_{1})\cdots A_{\mu_{n}}^{a_{n}}(x_{n})\big\rangle,
\end{equation}
where,
\begin{equation}\label{eq:measure}
\left[d^{5}x_{i}\right]=\sqrt{-g(x_{i})}\sqrt{2E_{p_i}}d^{5}x_{i}
\end{equation}
and the field modes,
\begin{equation}\label{eq:gluon-mode}
f_{k\mu}^{a\,h}(x)=\,\frac{\varepsilon_{\mu}^{a,h}(k)}{\sqrt{2E_{k}}}\!\left(1+\frac{x^{2}}{8\ell^{2}}\right)e^{ik\cdot x}.
\end{equation}
are described in \S\ref{sec:gf}. We also have the derivative operator $\mathcal{D}_x$ defined in \eqref{eq:derivative}.

Using the LSZ formula \eqref{eq:lsz}, we can write compute the amplitude for diagram (a,b) in fig. \ref{fig:soft_gluon} as follows:

\begin{figure}[h]
\centering
\begin{tikzpicture}[line width=1.5pt, scale=1.4]
\begin{scope}[xshift=0cm]
\begin{feynman}
\begin{scope}[shift={(6,0)}]
\draw[gluon] (.7,0)--(2,0);
\draw[gluon][rotate=30] (.7,0)--(2,0);
\draw[gluon][rotate=180] (.7,0)--(1.5,0);
\draw[gluon][rotate=180] (2.5,0)--(1.5,0);
\draw[gluon][rotate=150] (.7,0)--(2,0);
\draw[gluon, darkred][rotate=180] (1.5,0)--(1.5,1.3);
\node at (-0.8,.7) {$\bullet$};
\node at (-0.2,.95) {$\bullet$};
\node at (0.5,.9) {$\bullet$};
\node at (-1.5,.2) {$z'$};
\node at (-.8,-.2) {$z$};
\node[darkred] at (-1.5,-1.5) {$(\varepsilon_{\mu }, k)$};
\node at (-2.5,.2) {$p_1$};
\begin{scope}[shift={(0,0)}, scale=2]
    \draw [ultra thick] (0,0) circle (.35);
    \clip (0,0) circle (.3cm);
    \foreach \x in {-.9,-.8,...,.3}
        \draw[line width=1 pt] (\x,-.3) -- (\x+.6,.3);
\end{scope}
\end{scope}
\end{feynman}
\end{scope}
\end{tikzpicture}
\label{fig:soft_gluon2}
\end{figure}
\begin{align}
&\mathcal{A}^{^{\text{YM}}\left(1\right)}(p_{1},\ldots,p_{n-1},k)=\int d^{5}x_{1}\,\sqrt{-g(x_{1})}\;f_{p_{1}\gamma}^{*\,a'_{1},h_{1}}(x_{1})\eta^{\gamma\mu'_{1}}\;(-i)\mathcal{D}_{x_{1}}\nn\\&\qquad\times\prod_{j=2}^{n-1}\int d^{5}x_{j}\,\sqrt{-g(x_{j})}\;f_{p_{j}\sigma}^{*\,a_{j},h_{j}}(x_{j})\delta_{\mu_{j}}^{\sigma}\;(-i)\mathcal{D}_{x_{j}}\int d^{5}y\sqrt{-g(y)}\ f_{k\delta}^{*\,bh}(y)\eta^{\delta\rho}\ (-i)\mathcal{D}_{y}\;\nn\\&\qquad\times\int d^{5}z'\int d^{5}z\sqrt{-g\left(z\right)}\;\big\langle\mathcal{T}A_{\mu'_{1}}^{a'_{1}}\left(x_{1}\right)A_{\rho}^{b}\left(y\right)\mathcal{L}_{3}^{^{\text{YM}}}\left(z'\right)A_{\mu_{1}}^{a_{1}}\left(z\right)\big\rangle G_{n-1}^{a_{1}\ldots a_{n-1};\mu_{1}\ldots\mu_{n-1}}\left(z,2,\ldots,n-1\right).
\end{align}
Similarly,
\begin{align}
&\mathcal{A}^{^{\text{YM}}\left(n-1\right)}(p_{1},\ldots,p_{n-1},k)=\int d^{5}x_{n-1}\,\sqrt{-g(x_{n-1})}\;f_{p_{n-1}\gamma}^{*\,a'_{n-1},h_{n-1}}(x_{n-1})\eta^{\gamma\mu'_{n-1}}\;(-i)\mathcal{D}_{x_{n-1}}\nn\\&\qquad\times\prod_{j=1}^{n-2}\int d^{5}x_{j}\,\sqrt{-g(x_{j})}\;f_{p_{j}\sigma}^{*\,a_{j},h_{j}}(x_{j})\delta_{\mu_{j}}^{\sigma}\;(-i)\mathcal{D}_{x_{j}}\int d^{5}y\sqrt{-g(y)}\ f_{k\delta}^{*\,bh}(y)\eta^{\delta\rho}\ (-i)\mathcal{D}_{y}\nn\\&\qquad\times\int d^{5}z'\int d^{5}z\sqrt{-g\left(z\right)}\big\langle\mathcal{T}A_{\mu'_{n-1}}^{a'_{n-1}}\left(x_{n-1}\right)A_{\rho}^{b}\left(y\right)\mathcal{L}_{3}^{^{\text{YM}}}\left(z'\right)A_{\mu_{n-1}}^{a_{n-1}}\left(z\right)\big\rangle G_{n-1}^{a_{1}\ldots a_{n-1};\mu_{1}\ldots\mu_{n-1}}\left(1,\ldots,n-2,z\right).
\end{align}
Here, $G_{n-1}$ is the remaining $\left(n-1\right)$ point amplitude (with one leg amputated) that receives contribution from all kinds of vertices.
We can now evaluate the time ordered correlators using Wick contraction. It will suffice to demonstrate the calculation for $\mathcal{A}^{^\text{YM}\left(1\right)}$,
\begin{align}\label{first}
&\big\langle\mathcal{T}A_{\mu'_{1}}^{a'_{1}}\left(x_{1}\right)A_{\rho}^{b}\left(y\right)\mathcal{L}_{3}^{^\text{YM}}\left(z'\right)A_{\mu_{1}}^{a_{1}}\left(z\right)\big\rangle\nn\\&=\sqrt{-g\left(z'\right)}\,\,\gym\,\,\,f^{adc}\,g^{\delta\mu}\left(z'\right)g^{\sigma\nu}\left(z'\right)\big\langle\mathcal{T}A_{\mu'_{1}}^{a'_{1}}\left(x_{1}\right)A_{\rho}^{b}\left(y\right)\left(\partial_{\mu}A_{\nu}^{a}\left(z'\right)\right)A_{\delta}^{d}\left(z'\right)A_{\sigma}^{c}\left(z'\right)A_{\mu_{1}}^{a_{1}}\left(z\right)\big\rangle\nn\\&=\sqrt{-g\left(z'\right)}\,\,\gym\,\,\,f^{adc}\,g^{\delta\mu}\left(z'\right)g^{\sigma\nu}\left(z'\right)\nn\\&\times\left[D_{\rho\delta}^{bd}(y,z')D_{\sigma\mu_{1}}^{ca_{1}}(z',z)\partial_{\mu}D_{\mu_{1}'\nu}^{a_{1}'a}(x_1,z')+D_{\delta\mu_{1}}^{da_{1}}(z',z)D_{\rho\sigma}^{bc}(y,z')\partial_{\mu}D_{\mu_{1}'\nu}^{a_{1}'a}(x_1,z')\right.\nn\\&\,\,\,\,\left.+D_{\mu_{1}'\delta}^{a_{1}'d}(x_1,z')D_{\sigma\mu_{1}}^{ca_{1}}(z',z)\partial_{\mu}D_{\rho\nu}^{ba}(y,z')+D_{\mu_{1}'\sigma}^{a_{1}'c}(x_1,z')D_{\delta\mu_{1}}^{da_{1}}(z',z)\partial_{\mu}D_{\rho\nu}^{ba}(y, z')\right.\nn\\&\,\,\,\,\left.+D_{\mu_{1}'\delta}^{a_{1}'d}(x_1,z')D_{\rho\sigma}^{bc}(y,z')\partial_{\mu}D_{\mu_{1}\nu}^{a_{1}a}(z',z)+D_{\mu_{1}'\sigma}^{a_{1}'c}(x_1,z')D_{\rho\delta}^{bd}(y,z')\partial_{\mu}D_{\mu_{1}\nu}^{a_{1}a}(z',z)\right]
\end{align}
Using this, the overall integral over $y$ and then $x_1$ yields,
\begin{align}
&\mathcal{A}^{^{\text{YM}}\left(1\right)}(p_{1},\ldots,p_{n-1},k)=\int d^{5}z'\int d^{5}z\,\sqrt{-g(z)}\,\sqrt{-g(z')}\,\,\gym\,f^{adc}\,g^{\alpha\mu}g^{\beta\nu}\nn\\&\qquad\qquad\times\Big[f_{k\alpha}^{*\,dh}(z')D_{\beta\mu_{1}}^{ca_{1}}(z',z)\partial'_{\mu}f_{p_{1}\nu}^{*\,a,h_{1}}(z')+f_{k\beta}^{*\,ch}(z')D_{\alpha\mu_{1}}^{da_{1}}(z',z)\partial'_{\mu}f_{p_{1}\nu}^{*\,a,h_{1}}(z')\nn\\&\qquad\qquad\;+f_{p_{1}\alpha}^{*\,d,h_{1}}(z')D_{\beta\mu_{1}}^{ca_{1}}(z',z)\partial'_{\mu}f_{k\nu}^{*\,ah}(z')+f_{p_{1}\beta}^{*\,c,h_{1}}(z')D_{\alpha\mu_{1}}^{da_{1}}(z',z)\partial'_{\mu}f_{k\nu}^{*\,ah}(z')\nn\\&\qquad\qquad\;+f_{k\beta}^{*\,ch}(z')f_{p_{1}\alpha}^{*\,d,h_{1}}(z')\partial'_{\mu}D_{\mu_{1}\nu}^{a_{1}a}(z',z)+f_{k\alpha}^{*\,dh}(z')f_{p_{1}\beta}^{*\,c,h_{1}}(z')\partial'_{\mu}D_{\mu_{1}\nu}^{a_{1}a}(z',z)\Big]\nn\\&\qquad\times\prod_{j=2}^{n-1}\int d^{5}x_{j}\,\sqrt{-g(x_{j})}\;f_{p_{j}\sigma}^{*\,a_{j},h_{j}}(x_{j})\delta_{\mu_{j}}^{\sigma}\;(-i)\mathcal{D}_{x_{j}}G_{n-1}^{a_{1}\ldots a_{n-1};\mu_{1}\ldots\mu_{n-1}}\left(z,2,\ldots,n-1\right).
\end{align}
Now we need to expand the amplitude to $\mathcal{O}\left(\ell^{-2}\right)$ in the de Sitter length scale. From \eqref{eq:gluon-mode} and \eqref{prop2} it follows that,
\begin{equation}
\partial_{\mu}f_{k\nu}^{*ah}\left(x\right)=\frac{\varepsilon_{\nu}^{ah}\left(k\right)}{\sqrt{2E_{k}}}\left[\frac{x_{\mu}}{4\ell^{2}}+\left(1+\frac{x^{2}}{8\ell^{2}}\right)ik_{\mu}\right]e^{ik\cdot x}
\end{equation}
\begin{align}
\partial_{\mu}^{'}D_{\nu'\nu}^{a'a}\left(z',z\right)=&-i\eta_{\nu'\nu}\delta^{a'a}\ \int\frac{d^{5}k}{(2\pi)^{5}}\frac{e^{ik\cdot\left(z'-z\right)}}{k^{2}-\frac{3}{4\ell^{2}}}\nn\\&\times\left[\left(1+\frac{z'^{2}+z^2}{8\ell^{2}}\right)ik_{\mu}+\frac{z'_{\mu}}{4\ell^{2}}\right].
\end{align}
and,
\begin{align}
\sqrt{-g(x)}\,\mathcal{D}_{x}\partial_{\rho}D_{\mu\nu}^{ab}\left(x,y\right)=&\eta_{\mu\nu}\delta^{ab}\left(1-\frac{x^{2}}{\ell^{2}}\right)\partial_{\rho}\delta^{(5)}\left(x-y\right)\nn\\&+\frac{1}{2\ell^{2}}\eta_{\mu\nu}\delta^{ab}x_{\rho}\delta^{(5)}\left(x-y\right)+\frac{1}{2\ell^{2}}\partial_{\rho}D_{\mu\nu}^{ab}\left(x,y\right).
\end{align}
We also expand the metric using \eqref{eq: metric_expand}. Then using these and simplifying we eventually get,
\begin{align}
&\mathcal{A}^{^{\text{YM}}\left(1\right)}(p_{1},\ldots,p_{n-1},k)=-i\,\gym\,f^{adc}\int d^{5}z\;\frac{e^{-i\,q\cdot z}}{q^{2}}\nn\\&\qquad\times\Biggl\{\delta^{ca_{1}}\epsilon_{\mu_{1}}^{ah_{1}}(p_{1})\Biggl[2i\left(p_{1}\cdot\epsilon^{dh}(k)\right)\Bigl(1-\frac{z^{2}}{\ell^{2}}-\frac{i\left(q\cdot z\right)}{2\ell^{2}q^{2}}+\frac{5}{4\ell^{2}q^{2}}\Bigr)+\frac{\left(z\cdot\epsilon^{dh}(k)\right)}{4\ell^{2}}\Biggr]\nn\\&\qquad\qquad+\delta^{da_{1}}\left\{ \epsilon^{ch}(k)\cdot\epsilon^{ah_{1}}(p_{1})\right\} \Biggl[i\left(p_{1}-k\right)_{\mu_{1}}\Bigl(1-\frac{z^{2}}{\ell^{2}}-\frac{i\left(q\cdot z\right)}{2\ell^{2}q^{2}}+\frac{1}{\ell^{2}q^{2}}\Bigr)\Biggr]\nn\\&\qquad\qquad+\delta^{ca_{1}}\epsilon_{\mu_{1}}^{ah}(k)\Biggl[2i\left(k\cdot\epsilon^{dh_{1}}(p_{1})\right)\Bigl(1-\frac{z^{2}}{\ell^{2}}-\frac{i\left(q\cdot z\right)}{2\ell^{2}q^{2}}+\frac{1}{\ell^{2}q^{2}}\Bigr)+\frac{\left(z\cdot\epsilon^{dh_{1}}(p_{1})\right)}{4\ell^{2}}+\frac{i\left(k\cdot\epsilon^{dh_{1}}(p_{1})\right)}{2\ell^{2}q^{2}}\Biggr]\Biggr\}\nn\\&\qquad\qquad\times i\prod_{j=2}^{n-1}\int d^{5}x_{j}\,\sqrt{-g(x_{j})}\,f_{p_{j}\sigma}^{*a_{j}h_{j}}(x_{j})\,\delta_{\mu_{j}}^{\sigma}(-i)\mathcal{D}_{x_{j}}G_{n-1}^{a_{1}\cdots a_{n-1};\mu_{1}\cdots\mu_{n-1}}(z,2,\ldots,n-1)\,.
\end{align}
where $q=-p_1-k$.

The next step is to expand around the soft limit. We write $k=\omega\,\hat{k}$. Then the soft limit is $\omega\rightarrow 0$. Recall that we are working in the limit where $\omega\gg1/\ell$, i.e. the wavelength of the soft particle is much smaller than the de Sitter length $\ell$. Therefore we introduce a dimensionless parameter $\delta=\omega\,\ell$, which we keep fixed while taking the soft limit.In our limit, $\delta\gg 1$. The de Sitter corrections are then given by an expansion in $1/\delta$. Expanding the amplitude to $\mathcal{O}\left(\omega^0\right)$ we have,
\begin{align}
&\mathcal{A}^{^{\text{YM}}\left(1\right)}(p_{1},\ldots,p_{n-1},k)=\,\gym\,f^{ada_{1}}\left[\epsilon_{\mu_{1}}^{ah_{1}}(p_{1})\frac{p_{1}\cdot\epsilon^{dh}(k)}{\omega\,p_{1}\cdot\hat{k}}+\epsilon_{\mu_{1}}^{ah}(k)\frac{\epsilon^{dh_{1}}(p_{1})\cdot\hat{k}}{p_{1}\cdot\hat{k}}\nn\right.\\&\quad\quad\quad\quad\left.+\epsilon_{\mu_{1}}^{ah_{1}}(p_{1})\left\{ \frac{p_{1}\cdot\epsilon^{dh}(k)}{p_{1}\cdot\hat{k}}\left(\hat{k}\cdot\partial_{p_{1}}\right)+\frac{1}{4}\frac{p_{1}\cdot\epsilon^{dh}(k)}{\delta^{2}\left(p_{1}\cdot\hat{k}\right)^{2}}\left(p_{1}\cdot\partial_{p_{1}}\right)-\frac{1}{8}\frac{p_{1}\cdot\epsilon^{dh}(k)}{\delta^{2}\left(p_{1}\cdot\hat{k}\right)^{2}}\right\} \right]\nn\\&\quad\quad\quad\quad\times i\int d^{5}z\;e^{ip_{1}\cdot z}\prod_{j=2}^{n-1}\int d^{5}x_{j}\,\sqrt{-g(x_{j})}\,f_{p_{j}\sigma}^{*a_{j}h_{j}}(x_{j})\,\delta_{\mu_{j}}^{\sigma}(-i)\mathcal{D}_{x_{j}}G_{n-1}^{a_{1}\cdots a_{n-1};\mu_{1}\cdots\mu_{n-1}}(z,2,\ldots,n-1)\,.
\end{align}
where we have shifted the $z$ integral to the right by substituting $z\rightarrow-i\,\partial_{p_1}$.
Doing the $z$ integral using \eqref{eq:lsz} we finally get,
\begin{align}
&\mathcal{A}^{^{\text{YM}}\left(1\right)}(p_{1},\ldots,p_{n-1},k)=\,\gym\,\left(\text{ad}_{T_{d}}\right)^{aa_{1}}\nn\\&\qquad\qquad\times\left[\epsilon_{\mu_{1}}^{ah_{1}}(p_{1})\frac{p_{1}\cdot\epsilon^{dh}(k)}{\omega\,p_{1}\cdot\hat{k}}+\epsilon_{\mu_{1}}^{ah_{1}}(p_{1})\frac{p_{1}\cdot\epsilon^{dh}(k)}{p_{1}\cdot\hat{k}}\left(\hat{k}\cdot\partial_{p_{1}}\right)+\epsilon_{\nu}^{ah_{1}}(p_{1})\frac{\epsilon_{\mu}^{dh}(k)\hat{k}_{\sigma}}{p_{1}\cdot\hat{k}}(S^{\mu\sigma})^{\nu}{}_{\mu_{1}}\right.\nn\\&\qquad\qquad\qquad\left.+\frac{1}{\delta^{2}}\epsilon_{\mu_{1}}^{ah_{1}}(p_{1})\left\{ \frac{1}{4}\frac{p_{1}\cdot\epsilon^{dh}(k)}{\left(p_{1}\cdot\hat{k}\right)^{2}}\left(p_{1}\cdot\partial_{p_{1}}\right)-\frac{1}{8}\frac{p_{1}\cdot\epsilon^{dh}(k)}{\left(p_{1}\cdot\hat{k}\right)^{2}}\right\} \right]\Gamma_{n-1}^{a_{1}\mu_{1}}\left(p_{1},\ldots,p_{n-1}\right)
\end{align}
where, $\Gamma_{n-1}\left(p_{1},\ldots,p_{n-1}\right)=\epsilon_{\mu_{1}}^{a_{1}h_{1}}(p_{1})\,\Gamma_{n-1}^{a_{1}\mu_{1}}\left(p_{1},\ldots,p_{n-1}\right)$. Also, we've rewritten the structure constants in terms of the $\mathfrak{su}$(N) generators $T_a$ in adjoint representation, $f^{abc}=i\left(\text{ad}_{T_{a}}\right)^{bc}$ and identified the spin angular momentum operator term in spin-1 (vector) representation,
\begin{equation}\label{eq:spin}
(S^{\mu\sigma})^{\nu}_{\rho}=\left(\eta^{\mu\nu}\delta_{\;\rho}^{\sigma}-\eta^{\sigma\nu}\delta_{\;\rho}^{\mu}\right).    
\end{equation}
Similar expression is there for the $\left(n-1\right)$ leg with $p_1\rightarrow p_{n-1}$.
To this then we need to add the contribution from the diagram where the gluon attaches to the internal line. This can be obtained using the standard way of imposing gauge invariance on the amplitude \cite{Bern}. The result is,
\begin{align}
\mathcal{A}^{^{\text{YM}}\left(\text{int}\right)}(p_{1},\ldots,p_{n-1},k)=&\,-\gym\,f^{ada_{1}}\epsilon_{\mu_{1}}^{ah_{1}}(p_{1})\epsilon^{dh}(k)\cdot\partial_{p_{1}}\Gamma_{n-1}^{a_{1}\mu_{1}}\left(p_{1},\ldots,p_{n-1}\right)\nn\\&-\gym\,f^{ada_{n-1}}\epsilon_{\mu_{n-1}}^{ah_{n-1}}(p_{n-1})\epsilon^{dh}(k)\cdot\partial_{p_{n-1}}\Gamma_{n-1}^{a_{n-1}\mu_{n-1}}\left(p_{1},\ldots,p_{n-1}\right).
\end{align}
This term combines with the others to give us the soft gluon theorem with de Sitter corrections,
\begin{align}
&\mathcal{A}^{^{\text{YM}}}(p_{1},\ldots,p_{n-1},k)=\,\gym\,\left(\text{ad}_{T_{d}}\right)^{aa_{1}}\nn\\&\qquad\qquad\times\left[\epsilon_{\mu_{1}}^{ah_{1}}(p_{1})\frac{p_{1}\cdot\epsilon^{dh}(k)}{\omega\,p_{1}\cdot\hat{k}}+\epsilon_{\mu_{1}}^{ah_{1}}(p_{1})\frac{\epsilon_{\mu}^{dh}(k)\hat{k}_{\sigma}}{p_{1}\cdot\hat{k}}L_{\left(1\right)}^{\mu\sigma}+\epsilon_{\nu}^{ah_{1}}(p_{1})\frac{\epsilon_{\mu}^{dh}(k)\hat{k}_{\sigma}}{p_{1}\cdot\hat{k}}(S^{\mu\sigma})^{\nu}{}_{\mu_{1}}\right.\nn\\&\qquad\qquad\qquad\left.+\frac{1}{\delta^{2}}\epsilon_{\mu_{1}}^{ah_{1}}(p_{1})\left\{ \frac{1}{4}\frac{p_{1}\cdot\epsilon^{dh}(k)}{\left(p_{1}\cdot\hat{k}\right)^{2}}\left(p_{1}\cdot\partial_{p_{1}}\right)-\frac{1}{8}\frac{p_{1}\cdot\epsilon^{dh}(k)}{\left(p_{1}\cdot\hat{k}\right)^{2}}\right\} \right]\Gamma_{n-1}^{a_{1}\mu_{1}}\left(p_{1},\ldots,p_{n-1}\right)\nn\\&\qquad\qquad+\left\{ 1\rightarrow\left(n-1\right)\right\} ,
\end{align}
where $L^{\mu\nu}$ is the orbital angular momentum operator,
\begin{equation}
L_{\left(i\right)}^{\mu\nu}=\left(p_{i}^{\mu}\partial_{p_{i}}^{\nu}-p_{i}^{\nu}\partial_{p_{i}}^{\mu}\right).
\end{equation}
The structure of the curvature correction terms is similar to the massless scalar QED case discussed in \cite{Pratik}.

\section{Chern--Simons correction to the perturbative soft gluon theorem in de Sitter}\label{sec:CS}

Now let us add a Chern--Simons term to the usual gauge invariant action for QCD or QED. For demonstration, let us again consider an $n$ point colour ordered gluon amplitude $\Gamma_{n}(p_{1},\ldots,p_{n-1},k)$, where the $n$th gluon goes soft and hence $p_{n}=k$. But now we are working in a general Yang--Mills matter gauge theory with a Chern--Simons deformation. Then, as argued in \cite{Avi}, the leading Chern--Simons contribution (subleading in soft momenta overall $\mathcal{O}\left(\omega^0\right)$) comes solely from the diagram where the soft gluon attaches to $p_{1}$ or $p_{n-1}$ (see the diagram in the left in fig. \ref{fig:soft_gluon}) ,
\begin{equation}
\mathcal{A}^{\text{CS}}=\mathcal{A}^{\text{CS}\left(1\right)}+\mathcal{A}^{\text{CS}\left(n-1\right)}
\end{equation}
We calculate these two parts separately. This should then be added to the usual soft theorem with de Sitter corrections,
\begin{align}
\Gamma_{n}(\{p_{i}\},\omega\hat{k})=&\,\mathcal{A}^{\text{gauge}}+\mathcal{A}^{\text{CS}}\nn\\=&\left[\frac{1}{\omega}\mathbb{S}^{\left(0\right)}+\mathbb{S}^{\left(1\right)}+\frac{1}{\delta^{2}}\mathbb{S}'^{\left(1\right)}\right]\Gamma_{n-1}(\{p_{i}\})+\mathcal{A}^{\text{CS}}
\end{align}
where $\mathbb{S}^{\left(0\right)}$, $\mathbb{S}^{\left(1\right)}$ are the usual leading and subleading gauge soft factors and $\mathbb{S}'^{\left(1\right)}$ is the $1/\delta^2$ de Sitter correction,
\begin{align}
\mathbb{S}^{\left(0\right)}=&\,\gym\,\frac{p_{1}\cdot\epsilon(k)}{\,p_{1}\cdot\hat{k}}+\left\{1\rightarrow\left(n-1\right)\right\} \\\mathbb{S}^{\left(1\right)}=&\,\gym\,\frac{\epsilon_{\mu}(k)\hat{k}_{\sigma}}{p_{1}\cdot\hat{k}}J_{1}^{\mu\sigma}+\left\{ 1\rightarrow\left(n-1\right)\right\} \\\mathbb{S}'^{\left(1\right)}=&\,\gym\,\left\{ \frac{1}{4}\frac{p_{1}\cdot\epsilon(k)}{\left(p_{1}\cdot\hat{k}\right)^{2}}\left(p_{1}\cdot\partial_{p_{1}}\right)-\frac{1}{8}\frac{p_{1}\cdot\epsilon(k)}{\left(p_{1}\cdot\hat{k}\right)^{2}}\right\} +\left\{ 1\rightarrow\left(n-1\right)\right\} 
\end{align}
where now we've defined $\epsilon\left(k\right)=T_a\epsilon_a$ with the $\mathfrak{su}$(N) generators $T_a$ in the appropriate representation. Also, $J^{\mu\nu}=L^{\mu\nu}+S^{\mu\nu}$. Now let's calculate the Chern--Simons correction. The term in the action corresponding to the 3-point vertex is given by,
\begin{equation}
\mathcal{S}_{3}=\int d^{5}x\,\mathcal{L}_{3}.
\end{equation}
with
\begin{equation}
\mathcal{L}_{3}=\mathcal{L}_{3}^{^\text{gauge}}+\mathcal{L}_{3}^{^\text{CS}}
\end{equation}
where $\mathcal{L}_{3}^{^\text{gauge}}$ is the usual gauge-invariant action for QED or QCD and,
\begin{equation}
\mathcal{L}_{3}^{^\text{CS}}=\kappa\,\epsilon^{\mu\nu\rho\sigma\delta}A_{\mu}^{a}\partial_{\nu}A_{\rho}^{b}\partial_{\sigma}A_{\delta}^{c}\;\text{Tr}\left\{ T_{a}T_{b}T_{c}\right\}
\end{equation}
As in the previous section, again use the LSZ formula \eqref{eq:lsz} for $n$ gluon amplitude $\Gamma_n(p_{1},\ldots,p_{n})$ to compute the Chern--Simons correction,
\begin{figure}[h]
\centering
\begin{tikzpicture}[line width=1.5pt, scale=1.4]
\begin{scope}[xshift=0cm]
\begin{feynman}
\begin{scope}[shift={(6,0)}]
\draw[gluon] (.7,0)--(2,0);
\draw[gluon][rotate=30] (.7,0)--(2,0);
\draw[gluon][rotate=180] (.7,0)--(1.5,0);
\draw[gluon][rotate=180] (2.5,0)--(1.5,0);
\draw[gluon][rotate=150] (.7,0)--(2,0);
\draw[gluon, darkred][rotate=180] (1.5,0)--(1.5,1.3);
\node at (-0.8,.7) {$\bullet$};
\node at (-0.2,.95) {$\bullet$};
\node at (0.5,.9) {$\bullet$};
\node at (-1.5,.2) {$z'$};
\node at (-.8,-.2) {$z$};
\node[darkred] at (-1.5,-1.5) {$(\varepsilon_{\mu }, k)$};
\node at (-2.5,.2) {$p_1$};
\begin{scope}[shift={(0,0)}, scale=2]
    \draw [ultra thick] (0,0) circle (.35);
    \clip (0,0) circle (.3cm);
    \foreach \x in {-.9,-.8,...,.3}
        \draw[line width=1 pt] (\x,-.3) -- (\x+.6,.3);
\end{scope}
\end{scope}
\end{feynman}
\end{scope}
\end{tikzpicture}
\end{figure}
\begin{align}
&\mathcal{A}^{\text{CS}\left(1\right)}(p_{1},\ldots,p_{n-1},k)=\int d^{5}x_{1}\,\sqrt{-g(x_{1})}\;f_{p_{1}\gamma}^{*\,a'_{1},h_{1}}(x_{1})\eta^{\gamma\mu'_{1}}\;(-i)\mathcal{D}_{x_{1}}\nn\\&\qquad\times\prod_{j=2}^{n-1}\int d^{5}x_{j}\,\sqrt{-g(x_{j})}\;f_{p_{j}\sigma}^{*\,a_{j},h_{j}}(x_{j})\delta_{\mu_{j}}^{\sigma}\;(-i)\mathcal{D}_{x_{j}}\int d^{5}y\sqrt{-g(y)}\ f_{k\delta}^{*\,bh}(y)\eta^{\delta\rho}\ (-i)\mathcal{D}_{y}\nn\\&\qquad\times\int d^{5}z'\int d^{5}z\sqrt{-g\left(z\right)}\;\big\langle\mathcal{T}A_{\mu'_{1}}^{a'_{1}}\left(x_{1}\right)A_{\rho}^{b}\left(y\right)\mathcal{L}_{3}^{\text{CS}}\left(z'\right)A_{\mu_{1}}^{a_{1}}\left(z\right)\big\rangle G_{n-1}^{a_{1}\ldots a_{n-1};\mu_{1}\ldots\mu_{n-1}}\left(z,2,\ldots,n-1\right).
\end{align}
Similarly for $\left(n-1\right)$ vertex,
\begin{align}
&\mathcal{A}^{\text{CS}\left(n-1\right)}(p_{1},\ldots,p_{n-1},k)=\int d^{5}x_{n-1}\,\sqrt{-g(x_{n-1})}\;f_{p_{n-1}\gamma}^{*\,a'_{n-1},h_{n-1}}(x_{n-1})\eta^{\gamma\mu'_{n-1}}\;(-i)\mathcal{D}_{x_{n-1}}\nn\\&\times\prod_{j=1}^{n-2}\int d^{5}x_{j}\,\sqrt{-g(x_{j})}\;f_{p_{j}\sigma}^{*\,a_{j},h_{j}}(x_{j})\delta_{\mu_{j}}^{\sigma}\;(-i)\mathcal{D}_{x_{j}}\int d^{5}y\sqrt{-g(y)}\ f_{k\delta}^{*\,bh}(y)\eta^{\delta\rho}\ (-i)\mathcal{D}_{y}\nn\\&\times\int d^{5}z'\int d^{5}z\sqrt{-g\left(z\right)}\;\big\langle\mathcal{T}A_{\mu'_{n-1}}^{a'_{n-1}}\left(x_{n-1}\right)A_{\rho}^{b}\left(y\right)\mathcal{L}_{3}^{\text{CS}}\left(z'\right)A_{\mu_{n-1}}^{a_{n-1}}\left(z\right)\big\rangle G_{n-1}^{a_{1}\ldots a_{n-1};\mu_{1}\ldots\mu_{n-1}}\left(1,\ldots,n-2,z\right).
\end{align}
Again evaluating the time ordered correlators using Wick's theorem,
\begin{align}
&\big\langle\mathcal{T}A_{\mu'_{1}}^{a'_{1}}\left(x_{1}\right)A_{\rho}^{b}\left(y\right)\mathcal{L}_{3}^{\text{CS}}\left(z'\right)A_{\mu_{1}}^{a_{1}}\left(z\right)\big\rangle=\nn\\&\qquad\kappa\,\epsilon^{\mu\nu\eta\sigma\delta}\text{Tr}\left\{ T_{a}T_{d}T_{c}\right\} \big\langle\mathcal{T}A_{\mu'_{1}}^{a'_{1}}\left(x_{1}\right)A_{\rho}^{b}\left(y\right)A_{\mu}^{a}\left(z'\right)\partial_{\nu}A_{\eta}^{d}\left(z'\right)\partial_{\sigma}A_{\delta}^{c}\left(z'\right)A_{\mu_{1}}^{a_{1}}\left(z\right)\big\rangle\nn\\&\qquad=\kappa\,\epsilon^{\mu\nu\eta\beta\alpha}\text{Tr}\left\{ T_{a}T_{d}T_{c}\right\} \nn\\&\qquad\times\left[D_{\mu_{1}'\mu}^{a_{1}'a}(x_{1},z')\partial_{\nu}D_{\rho\eta}^{bd}(y,z')\partial_{\beta}D_{\alpha\mu_{1}}^{ca_{1}}(z',z)+D_{\mu_{1}'\mu}^{a_{1}'a}(x_{1},z')\partial_{\beta}D_{\rho\alpha}^{bc}(y,z')\partial_{\nu}D_{\eta\mu_{1}}^{da_{1}}(z',z)\right.\nn\\&\,\qquad\left.+D_{\rho\mu}^{ba}(y,z')\partial_{\nu}D_{\mu_{1}'\eta}^{a_{1}'d}(x_{1},z')\partial_{\beta}D_{\alpha\mu_{1}}^{ca_{1}}(z',z)+D_{\rho\mu}^{ba}(y,z')\partial_{\beta}D_{\mu_{1}'\alpha}^{a_{1}'c}(x_{1},z')\partial_{\nu}D_{\eta\mu_{1}}^{da_{1}}(z',z))\right.\nn\\&\,\qquad\left.+D_{\mu\mu_{1}}^{aa_{1}}(z',z)\partial_{\nu}D_{\mu_{1}'\eta}^{a_{1}'d}(x_{1},z')\partial_{\beta}D_{\rho\alpha}^{bc}(y,z')+D_{\mu\mu_{1}}^{aa_{1}}(z',z)\partial_{\beta}D_{\mu_{1}'\alpha}^{a_{1}'c}(x_{1},z')\partial_{\nu}D_{\rho\eta}^{bd}(y,z')\right].
\end{align}
Substituting this back we can combine various terms to get,
\begin{align}
\big\langle\mathcal{T}A_{\mu_{1}'}^{a_{1}'}(x_{1})A_{\rho}^{b}(y)\mathcal{L}_{3}^{\mathrm{CS}}(z')A_{\mu_{1}}^{a_{1}}(z)\big\rangle=&\frac{i\kappa}{2}\,\epsilon^{\mu\nu\eta\beta\alpha}d^{adc}\Big[D_{\mu_{1}'\mu}^{a_{1}'a}(x_{1},z')\,\partial_{\nu}D_{\rho\eta}^{bd}(y,z')\,\partial_{\beta}D_{\alpha\mu_{1}}^{ca_{1}}(z',z)\nonumber\\&\qquad\quad+D_{\rho\mu}^{ba}(y,z')\,\partial_{\nu}D_{\mu_{1}'\eta}^{a_{1}'d}(x_{1},z')\,\partial_{\beta}D_{\alpha\mu_{1}}^{ca_{1}}(z',z)\nonumber\\&\qquad\quad+D_{\mu\mu_{1}}^{aa_{1}}(z',z)\,\partial_{\nu}D_{\mu_{1}'\eta}^{a_{1}'d}(x_{1},z')\,\partial_{\beta}D_{\rho\alpha}^{bc}(y,z')\Big].
\end{align}
we have simplified using $\frac{i}{2}d^{abc}=\text{Tr}\left[T^a\left\{T^b,T^c\right\}\right]$.
Expanding the expression till $\mathcal{O}\left(\ell^{-2}\right)$ and simplifying we get,
\begin{align}\label{eq:CS expanded amplitude}
&\mathcal{A}^{\text{CS}\left(1\right)}(p_{1},\ldots,p_{n-1},k)=\,\frac{\kappa}{2}\,\epsilon^{\mu\nu\eta\beta\alpha}\,d^{adc}\,\int d^{5}z\,\frac{e^{-iq\cdot z}}{q^{2}}\nn\\&\qquad\times\Biggl\{\eta_{\alpha\mu_{1}}\delta^{ca_{1}}\epsilon_{\mu}^{ah_{1}}(p_{1})\epsilon_{\eta}^{dh}(k)\left[k_{\nu}p_{1\beta}-\frac{ik_{\nu}z_{\beta}}{2\ell^{2}}-\frac{3i\,k_{\nu}p_{1\beta}\,(q\cdot z)}{2\ell^{2}q^{2}}-\frac{3z^{2}k_{\nu}p_{1\beta}}{4\ell^{2}}+\frac{5k_{\nu}p_{1\beta}}{2\ell^{2}q^{2}}+\frac{iq_{\beta}z_{\nu}}{4\ell^{2}}\right]\nn\\&\qquad+\eta_{\alpha\mu_{1}}\delta^{ca_{1}}\epsilon_{\mu}^{ah_{1}}(p_{1})\epsilon_{\eta}^{dh}(k)\left[k_{\nu}p_{1\beta}-\frac{ip_{1\beta}z_{\nu}}{2\ell^{2}}-\frac{3i\,p_{1\beta}k_{\nu}\,(q\cdot z)}{2\ell^{2}q^{2}}-\frac{3z^{2}p_{1\beta}k_{\nu}}{4\ell^{2}}+\frac{5p_{1\beta}k_{\nu}}{2\ell^{2}q^{2}}+\frac{iq_{\nu}z_{\beta}}{4\ell^{2}}\right]\nn\\&\qquad+\eta_{\alpha\mu_{1}}\delta^{ca_{1}}\epsilon_{\mu}^{ah_{1}}(p_{1})\epsilon_{\eta}^{dh}(k)\left[k_{\nu}p_{1\beta}-\frac{ip_{1\beta}z_{\nu}}{4\ell^{2}}-\frac{3i\,p_{1\beta}k_{\nu}\,(q\cdot z)}{2\ell^{2}q^{2}}-\frac{3z^{2}p_{1\beta}k_{\nu}}{4\ell^{2}}+\frac{5p_{1\beta}k_{\nu}}{2\ell^{2}q^{2}}-\frac{ik_{\nu}z_{\beta}}{4\ell^{2}}\right]\Biggr\}\nn\\&\qquad\times\prod_{j=2}^{n-1}\int d^{5}x_{j}\,\sqrt{-g(x_{j})}\,f_{p_{j}\sigma}^{*\,a_{j},h_{j}}(x_{j})\,\delta_{\mu_{j}}^{\sigma}(-i)\mathcal{D}_{x_{j}}G_{n-1}^{a_{1}\cdots a_{n-1};\mu_{1}\cdots\mu_{n-1}}(z,2,\ldots,n-1)\,.
\end{align}
where again $q=-p_1-k$ and Further expanding around the soft limit $\omega\rightarrow 0$ with  $\delta=\omega\,\ell$ held fixed we get,
\begin{align}
&\mathcal{A}^{\text{CS}\left(1\right)}(p_{1},\ldots,p_{n-1},k)=\frac{3\kappa}{4}\,\epsilon^{\mu\nu\eta\beta\alpha}d^{adc}\,\frac{\hat{k}_{\nu}p_{1\beta}}{p_{1}\cdot\hat{k}}\,\eta_{\alpha\mu_{1}}\,\delta^{ca_{1}}\,\epsilon_{\mu}^{ah_{1}}(p_{1})\epsilon_{\eta}^{dh}(k)\nn\\&\qquad\times\int d^{5}z\,e^{i\,p_{1}\cdot z}\,\prod_{j=2}^{n-1}\int d^{5}x_{j}\,\sqrt{-g(x_{j})}\,f_{p_{j}\sigma}^{*\,a_{j},h_{j}}(x_{j})\,\delta_{\mu_{j}}^{\sigma}(-i)\mathcal{D}_{x_{j}}G_{n-1}^{a_{1}\cdots a_{n-1};\mu_{1}\cdots\mu_{n-1}}(z,2,\ldots,n-1)\,.
\end{align}
Note that the $1/\ell^2$ terms in equation \eqref{eq:CS expanded amplitude} are $\mathcal{O}(\omega^2)$ and the $1/q^2\ell^2$ terms are $\mathcal{O}(\omega)$, therefore they both drop out from the subleading soft theorem. They do not contribute at $\mathcal{O}(\omega^0)$ but do contribute at $\mathcal{O}(\omega)$ or higher. Similarly for the $\left(n-1\right)$ leg,
\begin{align}
&\mathcal{A}^{\text{CS}\left(n-1\right)}(p_{1},\ldots,p_{n-1},k)=\,\frac{3\kappa}{4}\,\epsilon^{\mu\nu\eta\beta\alpha}d^{adc}\,\frac{\hat{k}_{\nu}p_{n-1;\beta}}{p_{n-1}\cdot\hat{k}}\,\eta_{\alpha\mu_{n-1}}\,\delta^{ca_{n-1}}\,\epsilon_{\mu}^{ah_{n-1}}(p_{n-1})\epsilon_{\eta}^{dh}(k)\nn\\&\qquad\times\int d^{5}z\,e^{i\,p_{n-1}\cdot z}\,\prod_{j=1}^{n-2}\int d^{5}x_{j}\,\sqrt{-g(x_{j})}\,f_{p_{j}\sigma}^{*\,a_{j},h_{j}}(x_{j})\,\delta_{\mu_{j}}^{\sigma}(-i)\mathcal{D}_{x_{j}}G_{n-1}^{a_{1}\cdots a_{n-1};\mu_{1}\cdots\mu_{n-1}}(1,\ldots,n-2,z).
\end{align}
Doing the final $z$ integral we get,
\begin{align}
\mathcal{A}^{\text{CS}}=\mathbb{S}^{\left(1\right)\text{CS}}\,\Gamma_{n-1}\left(p_{1},\ldots,p_{n-1}\right)
\end{align}
where $\mathbb{S}^{\left(1\right)\text{CS}}$ has the same form as in \cite{Avi},
\begin{equation}
\mathbb{S}^{\left(1\right)\text{CS}}\,\Gamma_{n-1}\left(p_{1},\ldots,p_{n-1}\right)=\widetilde{\sum_{j}}\frac{3\kappa}{4}\,\epsilon^{\mu\nu\eta\beta\alpha}d^{ada_{j}}\,\frac{\epsilon_{\eta}^{dh}(k)\hat{k}_{\nu}}{p_{j}\cdot\hat{k}}\,\eta_{\alpha\mu_{j}}\,\epsilon_{\mu}^{ah_{j}}(p_{j})\,p_{j\beta}\;\Gamma_{n-1}^{a_{j}\mu_{j}}\left(p_{1},\ldots,p_{n-1}\right).
\end{equation}
$\widetilde{\sum_{j}}$ sums over all external gluon legs in accordance with colour ordering, here $j=1\ \text{or}\ n-1$.

\section{Conclusion and future directions}

We showed that in de Sitter, the subleading soft theorems of Chern--Simons gauge theories have the form,
\begin{align}
\Gamma_{n}(\{p_{i}\},\omega\hat{k})=&\,\mathcal{A}^{\text{gauge}}+\mathcal{A}^{\text{CS}}\nn\\=&\left[\frac{1}{\omega}\mathbb{S}^{\left(0\right)}+\mathbb{S}^{\left(1\right)}+\frac{1}{\delta^{2}}\mathbb{S}'^{\left(1\right)}+\mathbb{S}^{\left(1\right)\text{CS}}\right]\Gamma_{n-1}(\{p_{i}\}).
\end{align}
where the $1/\delta^2$ de Sitter correction $\mathbb{S}'^{\left(1\right)}$ does not receive any contribution from the Chern--Simons term. Equivalently, the subleading Chern--Simons soft factor, $\mathbb{S}^{\left(1\right)\text{CS}}$ does not receive any $1/\delta^2$ correction.

It is suggestive from the analysis that higher order de Sitter corrections (i.e. $\mathcal{O}\left(1/\ell^4\right)$ or higher) will not correct the Chern--Simons soft factor at the subleading $\mathcal{O}\left(\omega^0\right)$ order. This is because every power of $1/\ell=\omega/\delta$ brings in a power of $\omega$ in the numerator and hence must be compensated by a factor of $1/q^2\propto1/\omega$ for it to contribute at subleading order $\mathcal{O}(\omega^0)$ in the soft expansion. But it is clear from the structure of the amplitude in the LSZ construction and curvature expansion of the field modes and the propagator that the higher curvature corrections (like $1/\ell^4$, $1/\ell^6$ etc.) do not come with adequate powers of $1/q^2$ to contribute to subleading order in the soft expansion. In other words, the higher order curvature corrections have the schematic form:
\begin{equation}
T_{1}\sim\frac{1}{\ell^{2n}}P_1(q)
\end{equation}
or
\begin{equation}
T_{2}\sim\frac{1}{q^{2n}\ell^{2n}}P_2(q)
\end{equation}
where $n$ is a positive integer and $P_{1,2}$ are polynomials in $q$. Now since $q^2\sim\mathcal{O}(\omega)$ and $\ell=\frac {\delta}{\omega}$, we have $T_1\sim\mathcal{O}(\omega^{2n})$ and $T_2\sim\mathcal{O}(\omega^{n})$. Since $n\geq1$, these terms will not correct the subleading soft factor. This is the same reason why there are no curvature corrections to the leading soft theorem in this expansion. This indicates that the Chern--Simons soft factors are completely independent of the perturbative de Sitter corrections. Therefore the topological nature of these terms in the action translates to a statement at the level of amplitudes-- the subleading Chern--Simons soft factor is independent of the dS metric perturbation. In this sense, these Chern--Simons soft factors are more "universal" than the subleading gauge soft factors, which receive $1/\ell^2$ corrections in de Sitter.

There are several related perturbative constructions of an AdS $S$-matrix in \cite{Banerjee:2024yiq}, \cite{Gadde:2022ghy} and \cite{Nenmeli:2026qaz}. These approaches are similar in spirit to ours in that scattering is effectively localized in a region whose size is small compared with the curvature radius, and curvature corrections are organized perturbatively in the inverse curvature radius. There are, however, important differences in the definition of the scattering observable and, in particular, in the treatment of massless external states.

The AdS $S$-matrix construction of the above references is formulated primarily for massive external particles. In this approach the flat-space limit is associated with the simultaneous large-$R$, large-$\Delta$ limit, where $R$ is the AdS length scale and $\Delta$ is the conformal dimension, with the physical mass ($m\sim\Delta/R$) held fixed. The momentum variables used there are conformally covariant embedding-space variables which reduce to ordinary flat-space momenta in the ($R\to\infty$) limit. At finite $R$, the breaking of ordinary translation invariance is reflected in derivatives of the momentum-conserving delta function, and the AdS $S$-matrix is identified with the coefficient of the undifferentiated delta function. The extension of the above prescriptions to genuinely massless external particles is therefore not immediate. 

More generally, the conventional notion of asymptotic out-states is not natural at finite AdS radius, since the timelike reflecting boundary causes outgoing radiation to return to the interaction region, giving rise to hereditary, non-local-in-time effects.

Our dS construction is somewhat different. We can directly define the massless propagators and modes of the gluon field using an ansatz which was developed in \cite{PD}. The perturbative $S$-matrix can also be defined consistently for massless external states. Also, in our case, the relevant hierarchy is
\begin{equation}
E\gg\omega\gg H=\ell^{-1},
\end{equation}
so that the emitted particle is soft relative to the hard scattering scale while its wavelength remains much smaller than the de Sitter curvature radius. The flat-space limit ($\ell\to\infty$) is therefore manifest within the perturbative construction, and the massless external states continuously reduce to the usual flat-space scattering states. In this sense, our calculation directly probes curvature corrections to massless soft theorems in the sub-curvature regime, rather than obtaining the massless sector as a limit of a massive AdS $S$-matrix construction.

In relation to Chern--Simons perturbations, an interesting future direction would be to investigate the asymptotic symmetries associated with these terms. In ordinary gauge theories, leading and subleading soft theorems can often be understood as Ward identities of asymptotic symmetries. Given the universal structure of the Chern--Simons soft factors found in \cite{Avi} and in the present work, it is natural to ask whether they admit a similar symmetry interpretation. In particular, it would be interesting to determine whether their universality can be traced to an underlying asymptotic symmetry, both in flat spacetime and in curved backgrounds such as de Sitter space.

\section*{Acknowledgments}
AW would like to thank R. Loganayagam, Ritwick Kumar Ghosh, Debanjan Karan, Ashik H. and Shridhar Vinayak for useful discussions. AW acknowledges the support of the Department of Atomic Energy, Government of India, under project no. RTI4019.
\appendix
\section{Gluon field in de Sitter spacetime}\label{sec:gf}
In this section, we find the mode expansion and propagator of the gluon field in de Sitter spacetime. We closely follow \cite{Sayali_Diksha}. The equation of motion of this field in stereographic coordinates to $O\left(\f{1}{\ell^2}\right)$ is given by,
\begin{align}\label{eq:wave-equation}
\left(1+\frac{x^2}{2\ell^2}\right)\left(\Box_x A^a_\mu-\p_\mu(\p^\nu A^a_\nu)\right)-\frac{1}{2\ell^2}[x.\p A^a_\mu -\p_\mu(x^\nu A_\nu)+A^a_\mu]=\ j^a_\mu,
\end{align}
where $j^a_\mu$ is some $\mathfrak{su}$(N) matter current and we have only kept $\mathcal{O}(\f{1}{\ell^2})$ corrections to flat space equation of motion. We can fix the gauge by setting \footnote{The most general gauge choice is illustrated in \cite{Sayali_Diksha}, which has the form $\partial^{\nu}A_{\nu}+\frac{(4-d)}{2l^2}x^{\nu}A_{\nu}=0$. This choice is engineered in a way such that the $x$-dependent terms coming from the equations of motion get eliminated resulting in some simplification. Further, this choice reduces to the standard Lorenz gauge in $d=4$ dimensions. Also, it has been shown previously in \cite{Sayali_Diksha},\cite{SCB} that the soft factors are indeed gauge invariant. This has been demonstrated in the soft photon and graviton cases and expected to remain true for gluons. Thus our results should not depend on this non covariant gauge fixing.}
\begin{align}\label{eq:gauge1}
\p^\nu A^a_\nu
-\f{1}{2\ell^2}\ (x^\nu A^a_\nu)=0.
\end{align}
or equivalently in momentum space,
\begin{align}\label{gauge}
ik^\nu\varepsilon^{ah}_\nu\  a^h_k\ e^{ik.x}
-\f{1}{4\ell^2}\varepsilon^{ah}_\mu x^\mu\ a^h_k\,e^{ik.x}=0 
\end{align}
Note that this gauge choice is not covariant. Using the gauge condition \eqref{eq:gauge1} in the equation of motion, we obtain
\begin{align}\label{eq:geom}
\Box_x A^a_\mu-\f{1}{2\ell^2}[x.\partial+1]A^a_\mu=j^a_\mu \left(1-\frac{x^2}{2\ell^2}\right).
\end{align}
The equations of motion \eqref{eq:geom} admit homogeneous solutions (i.e. $j^a_\mu = 0$) of the form 
\begin{align}\label{hk}
f^{ah}_{k\mu}(x) =\ \frac{1}{\sqrt{2 E_k}} \varepsilon_\mu^{ah}(k)\ \left(1+\f{1}{8}\frac{x^2}{\ell^2}\right)\ e^{ik.x},\ \ k^2=\f{3}{4\ell^2},
\end{align}
where $E_k$ is the zeroth component of $k^\mu$. After fixing all redundant degrees of freedom of the gluon field, one is left with $3$ physical degrees of freedom. Here we have used $h$ to denote these physical polarisation states with $h$ taking values from 1 to $3$. 

The inner product on the solution space is defined as
\begin{align}
(g^{\mu\nu}\ f^{ah}_{k\mu},f^{a'h'}_{k'\nu})=&-i\int d^{4}x\sqrt{-g}\ g^{\mu\nu}(x)\ e^{-\epsilon \f{|\vec{x}|}{R}}\ [\ f^{*ah}_{k\mu}(t,\vec{x})\ \p^t f^{a'h'}_{k'\nu}(t,\vec{x}) -f^{a'h'}_{k'\mu}(t,\vec{x})\ \p^t f^{*ah}_{k\nu}(t,\vec{x})\ ]\nn\\
=&\ {(2\pi)^{4}} \delta_{h,h'}\delta^{a'a} \delta^{(4)}(\vec{k}-\vec{k'}).\label{inA}
\end{align}
where we have again added exponential damping factor to get rid of the boundary terms. Hence, the gluon field can be expanded in terms of the above modes as follows
\begin{equation}\label{eq:mode-expansion}
A^a_\mu(x)=\sum_{h=1}^{3}\int\frac{d^{4}\vec{k}}{(2\pi)^{4}}[a^h_k\ f_{k\mu}^{ah}(x)+a^{h\dagger}_k \  f_{k\mu}^{*ah}(x)].
\end{equation}
Now let us look at the Feynman propagator of the gluon field. Again following the analysis for photons in \cite{Sayali_Diksha} we obtain,
\begin{equation}
\Big[\Box_x -\ \f{1}{2\ell^2}\ [x.\p +1]\ \Big] D^{aa'}_{\mu\nu'}(x,x') =i\eta_{\mu\nu'}\delta^{aa'}\delta^5(x-x')\left[1+\f{1}{4}\f{x^2}{\ell^2}\right].\label{photonG}
\end{equation}
This can also be rewritten by defining the derivative operator,
\begin{equation}\label{eq:derivative}
\mathcal{D}_x\equiv\left(1+\frac{x^2}{\ell^2}\right)\Box_x-\frac{1}{2\ell^2}\left(x\cdot\partial+1\right)
\end{equation}
such that now,
\begin{equation}\label{eq:prop-equation2}
\mathcal{D}_x D^{aa'}_{\mu\nu'}(x,x')=i\,\eta_{\mu\nu}\delta^{\left(5\right)}\left(x-x'\right)\left(1+\frac{5}{4}\frac{x^2}{\ell^2}\right)
\end{equation}
The solution to the propagator is given by 
\begin{equation}
D^{aa'}_{\mu\nu'}(x,y)=-i\,\eta_{\mu\nu'}\delta^{aa'}\int \f{d^5 k}{(2\pi)^5}\f{e^{ik.(x-y)}}{k^2}\Big[1+\f{1}{\ell^2k^2}-\f{1}{2}\frac{ik.y}{\ell^2k^2}+\f{1}{4}\frac{y^2}{\ell^2}\Big].
\end{equation}
The propagator also admits an equivalent symmetric form in variables $x$ and $y$ given by,
\begin{equation}\label{prop2}
D^{aa'}_{\mu\nu'}(x,y) =-i\eta_{\mu\nu}\ \delta^{aa'}\int \f{d^5k}{(2\pi)^5}\f{e^{ik.(x-y)}}{k^2-\f{3}{4\ell^2}} \Big[1+\f{x^2+y^2}{8\ell^2}\Big].
\end{equation}
One can use integration by parts to show that these two forms are equivalent.
\section{LSZ for gluons}\label{gaugeap}
In this Appendix, we derive the LSZ reduction formula for gluons. The gluon field expansion is given in \eqref{eq:mode-expansion}. It can be inverted using the orthogonality of modes using \eqref{inA} to obtain
\begin{align}
a^h_p=(g^{\mu\nu}\ f^{ah}_{p\mu},\ A^{a})
\end{align}
The above relation will now be used to derive the LSZ formula. Note that
\begin{align}
&a^h_p(T)-a^h_p(-T)=\int^T_{-T} dt\ \p_t a^h_p = (g^{\mu\nu}\ f^{ah}_{p\mu},\ A^a_{k\nu})\nn\\
&=-i\int dt d^{4}x \ e^{-\epsilon \f{|\vec{x}|}{R}}\ \p_t \left(\sqrt{-g}\ g^{\mu\nu}(x)\  [\ f^{*ah}_{p\mu}(t,\vec{x})\ \p^t A^a_{k\nu}(t,\vec{x}) -A^a_{k\mu}(t,\vec{x})\ \p^t f^{*ah}_{p\nu}(t,\vec{x})\ ]\right) \nn \\
&=-i\int dt d^{4}x\ e^{-\epsilon \f{|\vec{x}|}{R}}\ \p_\rho \left(\sqrt{-g}\ g^{\mu\nu}(x)\  [\ f^{*ah}_{p\mu}(t,\vec{x})\ \p^\rho A^a_{k\nu}(t,\vec{x}) -A^a_{k\mu}(t,\vec{x})\ \p^\rho f^{*ah}_{p\nu}(t,\vec{x})\ ]\right) 
\end{align}
where $\rho$ runs over all spacetime indices. In the second line, we have added the terms at the spatial boundary. After taking $\epsilon \rightarrow 0$ limit, the above expression can be evaluated to get
\begin{align}
a^h_p(T)-a^h_p(-T)&=-i\int d^5 x \sqrt{-g(x)}\eta^{\mu \nu} f^{*ah}_{p\nu} \left(\left(1+ \f{x^2}{\ell^2}\right)\Box \ -\ \f{1}{2\ell^2}\ [x.\p +1]\right)A^a_\mu (x). \nn\\ 
\end{align}
Using
\begin{equation}\label{Dx}
\mathcal{D}_x = \left(1+ \f{x^2}{\ell^2}\right)\Box \ -\ \f{1}{2\ell^2}\ [x.\p +1],    
\end{equation}
we get
\begin{align}\label{lszA}
a^h_p(T)-a^h_p(-T)&=-i\int d^5 x \sqrt{-g(x)}\eta^{\mu \nu} f^{*ah}_{p\nu} (\mathcal{D}_x) A^a_\mu (x) 
\end{align}
Using \eqref{photonG}, we find that $\mathcal{D}_x$ acts on Greens function as follows:
\begin{align}\label{G11}
\mathcal{D}_x G_{\mu,\nu} (x,x') &= i\eta_{\mu\nu}\delta^5(x-x')\left[1+\f{5}{4}\f{x^2}{\ell^2}\right] \nn\\
&=  i\eta_{\mu\nu}\f{\delta^5(x-x')}{\sqrt{-g(x)}}
\end{align}

\bibliography{references}
\bibliographystyle{jhepbibstyle}

\end{document}